\documentclass[%
 reprint,
 groupedaddress,
 nofootinbib,
 nobibnotes,
 amsmath,amssymb,
 aps,
 prl,
]{revtex4-1}
\usepackage{graphicx}% Include figure files
\usepackage{dcolumn}% Align table columns on decimal point
\usepackage{bm}% bold math
\usepackage{booktabs}
\usepackage{amsmath,amssymb,amsbsy,amstext, amsthm, amsfonts}
\usepackage{color}
\usepackage{slashed}
\usepackage[dvipsnames]{xcolor}
\usepackage{tikz}
\usepackage{dsfont}
\usepackage{verbatim}
\usepackage[utf8]{inputenc} 
\usepackage{ragged2e}
\usepackage{physics}
\usepackage[hidelinks]{hyperref}

\def\Vhrulefill{\leavevmode\leaders\hrule height 0.7ex depth \dimexpr0.4pt-0.7ex\hfill\kern0pt}

\begin{document}

\title{Asymmetric Reheating via Inverse Symmetry Breaking}
\author{Aurora Ireland}
\thanks{
\href{mailto:anireland@uchicago.edu}{anireland@uchicago.edu}, 
ORCID: \href{https://orcid.org/0000-0001-5393-0971}{0000-0001-5393-0971}}
\affiliation{Department of Physics, University of Chicago, Chicago, IL 60637, USA}
\author{Seth Koren}
\thanks{
\href{mailto:skoren@nd.edu}{skoren@nd.edu}, 
ORCID: \href{https://orcid.org/0000-0003-2409-4171}{0000-0003-2409-4171}\\}
\affiliation{Enrico Fermi Institute, University of Chicago, Chicago, IL 60637, USA}
\affiliation{Department of Physics and Astronomy, University of Notre Dame, South Bend, IN, 46556, USA}
%\affiliation{University of Chicago, Department of Physics, Chicago, IL 60637, USA}

\date{\today}

\begin{abstract}
Asymmetric reheating is a generic requirement for models of dark sectors with light species, but its implementation is usually in tension with unique phenomenologies otherwise possible in compelling theories containing dark copies of the Standard Model.
We present a simple module to implement asymmetric reheating during a $\mathbb{Z}_2$-breaking phase \emph{above} some critical temperature.
This reinvigorates the possibility of an exactly degenerate mirror sector and the striking phenomenology of composite particles oscillating into their mirror counterparts.
\end{abstract}

\maketitle

\makeatletter
\renewcommand*{\@fnsymbol}[1]{\ensuremath{\ifcase#1\or * \or \dagger \or \ddagger\or
   \mathsection\or \mathparagraph\or \|\or **\or \dagger\dagger
   \or \ddagger\ddagger \else\@ctrerr\fi}}
\makeatother

%%%%%%%%%%%%%%%%%%%%%%%%%%%%%%%%%%%%%%%%%%%%%%%%%%%%%%%%%%%%%%%%%%%%
%%%%%%%%%%%%%%%%%%%%%%%%%%%%%%%%%%%%%%%%%%%%%%%%%%%%%%%%%%%%%%%%%%%%
\section{Introduction}

The phenomenology of extended dark sectors is powerfully constrained by early universe data, particularly as dark sectors with light degrees of freedom may be probed through purely gravitational effects. Precision measurements of cosmological parameters such as $N_{\rm eff}$ significantly circumscribe particle physics models and have led to the generic need for some mechanism of `asymmetric reheating', whereby the dark sector is populated at a lower temperature \cite{Feng:2008mu,Ackerman:2008kmp,Foot:2014uba,Reece:2015lch,Adshead:2016xxj,Hardy:2017wkr,Tenkanen:2016jic,Adshead:2019uwj,Lozanov:2019jxc,Sandick:2021gew,Chiu:2022bni}. 

Particularly affected are mirror models, which introduce a $\mathbb{Z}_2$-symmetric copy of the Standard Model (SM) fields and gauge groups \cite{Lee:1956qn,Kobzarev:1966qya,Pavsic:1974rq,Blinnikov:1982eh,Foot:1991bp,Foot:1991py,Kolb:1985bf,Hodges:1993yb,Krauss:1985ac,Berezhiani:1995am,Okun:2006eb}. Known schemes for implementing asymmetric reheating in mirror models generally require a broken $\mathbb{Z}_2$ in the late universe \cite{Foot:2000tp,Berezhiani:2000gw,Ignatiev:2000yy,Berezhiani:2003xm,Barbieri:2005ri,Barbieri:2016zxn,Craig:2016lyx,Chacko:2016hvu,Csaki:2017spo,Chacko:2018vss,Curtin:2021alk,Beauchesne:2021opx}\footnote{See \cite{Cline:2021fdy,Babu:2021mjg} for recent, complementary work on asymmetric reheating with minimal symmetry breaking.}, which limits perhaps their most interesting observational signature: The oscillations of neutral SM particles into their mirror counterparts. This includes oscillations between SM and mirror neutrinos \cite{Foot:1993yp,Foot:1995pa,Berezhiani:1995yi,Silagadze:1995tr,Foot:1997jf,Collie:1998ty,Foot:1999ph,Blinnikov:1999ky} and photons \cite{Foot:2000vy,Foot:2000aj,Foot:2002iy,Badertscher:2003rk,Berezhiani:2005ek,Berezhiani:2008gi,Foot:2012ai,Chacko:2019jgi,Curtin:2019lhm,Curtin:2019ngc,Koren:2019iuv,Howe:2021neq,Alizzi:2021vyc}, as well as oscillations between entire composite particles in the limit of an exact $\mathbb{Z}_2$ symmetry.
%but an exact $\mathbb{Z}_2$ symmetry allows for a far more striking possibility: oscillations between entire composite particles. 
Oscillations between neutrons and mirror neutrons have seen much study in light of neutron lifetime anomalies \cite{Mohapatra:2005ng,Mohapatra:2017lqw,Babu:2021mjg,Berezhiani:2005hv,Berezhiani:2006je,Serebrov:2007gw,Sandin:2008db,Serebrov:2008her,Berezhiani:2009ldq,Berezhiani:2011da,Berezhiani:2012rq,Berezhiani:2017azg,Berezhiani:2018eds,Berezhiani:2018udo,Beradze:2019yyp,Berezhiani:2020vbe,Kamyshkov:2021kzi}. 
More exotically, oscillation of entire hydrogen atoms into their mirror forms has recently been shown to have interesting effects in late-time cosmology \cite{Johns:2020mmo,Johns:2020rtp}.

In the related Twin Higgs literature, the cosmological concerns are often dealt with by simply abandoning the full $\mathbb{Z}_2$ symmetry at the level of the spectrum
\cite{Craig:2015pha, Barbieri:2016zxn,Harigaya:2019shz}. However, this explicit breaking dramatically restricts the phenomenology of these models, and consequently mirror worlds have received less attention of late. With the aim of reviving these interesting possibilities, we seek to have a mirror sector which is exactly degenerate with our own.

Our tool is the richness of phase structures allowed in finite temperature quantum field theory, as first clearly demonstrated by Weinberg \cite{Weinberg:1974hy}. Counter-intuitively, it is possible to have a mirror symmetry which is broken only \textit{above} a critical temperature---a phenomenon known as `inverse symmetry breaking'. Scalar fields receive corrections to their mass from interactions with other particles in the thermal plasma, and negative cross-quartic interactions with other scalars yield negative contributions to the finite temperature mass. A scalar may then develop a vacuum expectation value (vev) at high temperature when the thermal contribution to its mass dominates \cite{Roos:1995vm,Orloff:1996yn,Bajc:1999cn,Pinto:2006cb}. Early concerns that such phenomena might be artifacts of fixed-order perturbation theory have been alleviated by follow-up work on the lattice, robustly evincing high-temperature symmetry-breaking phases \cite{Jansen:1998rj,Bimonte:1999tw,Pinto:1999pg}.

We present here a minimal module to implement the asymmetric reheating of a degenerate mirror sector via inverse symmetry breaking. The idea is to use the high-temperature $\mathbb{Z}_2$-breaking phase to set up an initial asymmetry in the energy densities of the SM and mirror sectors. The immediate model-building challenge is that we are asking for effects derived from thermal equilibrium to result in a far-out-of-equilibrium configuration. Further, this asymmetry in abundances must persist at late times once symmetry is restored and the sectors become exactly degenerate at low temperatures. 

\begin{figure}
    \centering
    \includegraphics[width=0.9\linewidth]{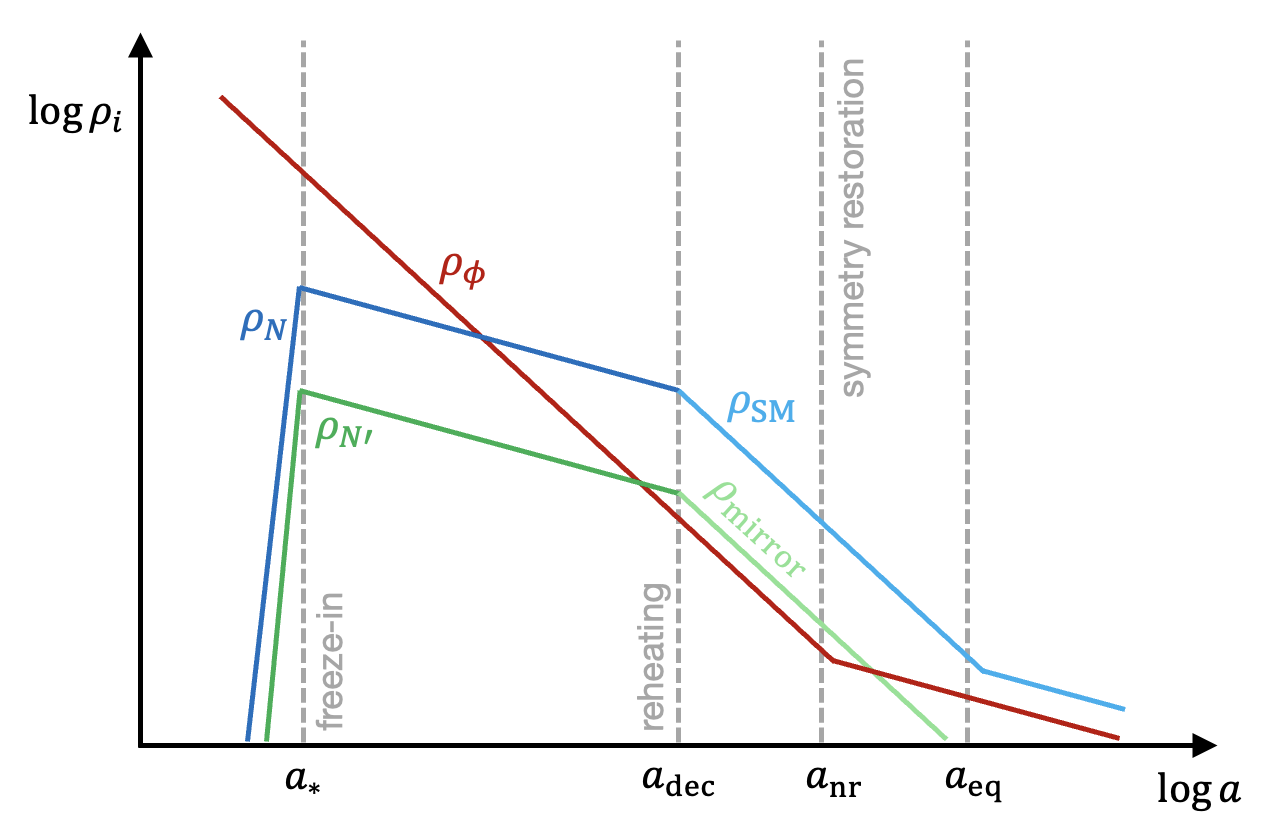}
    \caption{An overview of our cosmological timeline. At early times and high-$T$, the scalar sector dominates and the $\mathbb{Z}_2$ is spontaneously broken. At a value of the scale factor $a_*$ corresponding to $T_* \sim M_N$, freeze-in occurs, resulting in an asymmetric yield of heavy right-handed neutrinos $\rho_N \gg \rho_{N'}$. These come to dominate the universe's energy budget before decaying to asymmetrically reheat the SM and mirror sectors at some $a_{\rm dec}$ corresponding to $T_{\rm dec} \sim y_\nu \sqrt{M_N M_{\rm Pl}}$. The $\mathbb{Z}_2$ is restored and the $\phi$'s become non-relativistic at $a_{\rm nr}$, corresponding to $T \sim \mu_\pm$, after which they remain a component of the dark matter. See text for definitions and futher details.}
    \label{fig:timeline}
\end{figure}

Our strategy will be to use a non-thermal production mechanism---freeze-in \cite{McDonald:2001vt,Hall:2009bx,Bernal:2017kxu}---to populate the SM and mirror sectors. We introduce an auxiliary $\mathbb{Z}_2$-breaking sector which is feebly-coupled to heavy right-handed SM and mirror neutrinos $N, N'$, where primes denote mirror species. Annihilations of scalars yield asymmetric abundances due to asymmetric couplings in the $\mathbb{Z}_2$-broken phase. The heavy, non-relativistic $N$ and $N'$ are long-lived and act as reheatons, with the asymmetric number density leading to an asymmetry in reheating temperatures $T_{\rm SM} > T_{\rm mirror}$. See Fig.~\ref{fig:timeline} for a schematic timeline.

%%%%%%%%%%%%%%%%%%%%%%%%%%%%%%%%%%%%%%%%%%%%%%%%%%%%%%%%%%%%%%%%%%%%
%%%%%%%%%%%%%%%%%%%%%%%%%%%%%%%%%%%%%%%%%%%%%%%%%%%%%%%%%%%%%%%%%%%%
\section{A Minimal Module}

We consider a theory of three sectors: the SM supplemented with heavy right-handed neutrinos $N$, a mirror copy (whose species are denoted by primes), and a thermal sector of two real scalar singlets: $\phi_+$ and $\phi_-$. Under the $\mathbb{Z}_2$ symmetry which exchanges the particles of the SM and mirror sectors, $\phi_-$ is odd while $\phi_+$ is even. After inflation, the scalar sector is reheated to high temperatures and $\phi_-$ develops a negative thermal mass, breaking the $\mathbb{Z}_2$ symmetry. During this broken phase, $\phi_+$ serves to populate the $N$ and $N'$. 

We first review inverse symmetry breaking in a sector of just two scalars, but note that we can have more freedom in realizing this scenario with more fields. We then demonstrate how freeze-in production of heavy right-handed neutrinos during the broken phase can translate to an asymmetry in reheating temperatures.

%%%%%%%%%%%%%%%%%%%%%%

%%%%%%%%%%%%%%%%%%%%%%%%%%%%%%%%%%%%%%%%%%%%%%%%%%%%%%%%%%%%%%%%%%%%
%%%%%%%%%%%%%%%%%%%%%%%%%%%%%%%%%%%%%%%%%%%%%%%%%%%%%%%%%%%%%%%%%%%%
\subsection{Inverse Symmetry Breaking}

At tree-level, the potential for the scalar sector reads\footnote{Note that for simplicity we have ignored the cubic couplings, since these do not qualitatively affect the phase structure at high temperatures. Further, the assumption that they are negligibly small will be consistent with the naturalness expectations in our freeze-in model.}
\begin{equation}\label{treelevel}
    V_0 = \frac{\mu_+^2}{2} \phi_+^2 + \frac{\mu_-^2}{2} \phi_-^2 + \frac{\lambda_+}{4} \phi_+^4 + \frac{\lambda_-}{4} \phi_-^4 + \frac{\lambda_{\pm}}{4} \phi_+^2 \phi_-^2 \,.
\end{equation}
While the quartic self-couplings must be positive for the potential to be bounded from below, $\lambda_{\pm}$ may be negative provided
\begin{equation}
    \lambda_{\pm} > -2 \sqrt{\lambda_+ \lambda_-} \,.
\end{equation}
At 1-loop, the potential receives radiative corrections described by the zero-temperature Coleman-Weinberg potential $V_{\rm CW}$ and the 1-loop thermal potential $V_{T}^{1\text{-loop}}$,
\begin{equation}
    V_{\rm eff}(\phi_i, T) = V_0(\phi_i) + V_{\rm CW}(\phi_i) + V_T^{1\text{-loop}}(\phi_i, T) \,,
\end{equation}
where $i=\pm$ and $T$ denotes the temperature of the scalar sector. See e.g. \cite{Quiros:1999jp,Curtin:2016} for a review. The thermal potential dominates for our high temperature regime of interest, at least until new degrees of freedom come in at a scale $\Lambda$. Working to leading order in the high-temperature expansion, the quadratic terms from which we find the leading order contributions to the masses are
\begin{equation}
    V_T^{1\text{-loop}} \simeq \frac{T^2}{48} (6 \lambda_{+} + \lambda_{\pm}) \phi_+^2 + \frac{T^2}{48} (6 \lambda_{-} + \lambda_{\pm}) \phi_-^2 + ...
\end{equation}
Defining the coefficients
\begin{equation}
    c_+ = \frac{1}{24}(6 \lambda_+ + \lambda_{\pm}) \,, \,\,\, c_- = \frac{1}{24}(6 \lambda_- + \lambda_{\pm}) \,,
\end{equation}
the masses for $\phi_+$ and $\phi_-$ are
\begin{equation}
    M_+^2(T) = \mu_+^2 + c_+ T^2 \,, \,\,\, M_-^2(T) = \mu_-^2 + c_- T^2 \,.
\end{equation}

Examining these, it is apparent how phenomena\footnote{There has recently been much interest in the related phenomenon of `symmetry non-restoration'---where the zero-temperature mass is also negative---which may have applications for the electroweak phase transition and baryogenesis \cite{Meade:2018saz,Servant:2018,Glioti:2018roy,Bajc:2020gpa,Matsedonskyi:2020mlz,Matsedonskyi:2020kuy,Chao:2021xqv,Biekotter:2021ysx,Carena:2021,Bai:2021hfb}.} like inverse symmetry breaking can arise from a negative cross-quartic coupling. If $\lambda_{\pm} < 0$ and $|\lambda_{\pm}| > 6 \lambda_-$, then $c_-$ becomes negative and $\phi_-$ develops a negative thermal mass $M_-^2(T) < 0$ for sufficiently high temperatures---the hallmark of spontaneous symmetry breaking. Thus, at high temperatures the theory will be in the $\mathbb{Z}_2$-broken phase,
while at zero temperature the symmetry will be intact---an instance of inverse symmetry breaking.

The zero-temperature vacuum located at $(\phi_+, \phi_-) = (0, 0)$ is $\mathbb{Z}_2$-symmetric, but as the temperature is increased there is a phase transition at the critical point
\begin{equation}
    T_c = \sqrt{\frac{\mu_-^2}{|c_-|}} \,.
\end{equation}
Above this temperature, the theory enters into the broken phase as $\phi_-$ develops the temperature-dependent vacuum expectation value $\left\langle\phi_-(T)\right\rangle \equiv v_-(T)$, given at leading order in the high temperature expansion by
\begin{equation}
    v_-(T) = \sqrt{-\frac{1}{\lambda_-}(\mu_-^2 + c_- T^2)} \simeq \sqrt{\frac{|c_-|}{\lambda_-}} T  \,.
\end{equation}

Note that while the high temperature expansion suffices for our purposes, a more detailed study of the phase transition and precise predictions for quantities like the critical temperature would require going beyond the 1-loop approximation and performing thermal resummation of the effective potential, as the perturbative expansion breaks down in the infrared \cite{Parwani:1992,Arnold:1993,Quiros:1999jp,Curtin:2016}. These technicalities will not concern us here, as our purpose is not to study this sector in detail.

%%%%%%%%%%%%%%%%%%%%%%%%%%%%%%%%%%%%%%%%%%%%%%%%%%%%%%%%%%%%%%%%%%%%
%%%%%%%%%%%%%%%%%%%%%%%%%%%%%%%%%%%%%%%%%%%%%%%%%%%%%%%%%%%%%%%%%%%%
\subsection{Freeze-In Production}

We wish to take advantage of this high-temperature $\mathbb{Z}_2$-broken phase to establish an asymmetry in the energy densities of the SM and mirror sectors. The freeze-in mechanism is a natural candidate to accomplish this since it populates states which are never in equilibrium with the thermal sector. 

We will focus on the following two portal operators between the auxiliary scalar sector and heavy right-handed neutrinos $N$ and $N'$ of mass $M_N$
\begin{equation}
    - \mathcal{L}_N = \lambda \, \phi_+ \, (N N + N' N') + \frac{C}{\Lambda}\, \phi_+ \, \phi_- \, (N N - N' N') \,,
\end{equation}
where the dimension-5 operator may be generated by integrating out heavier fields at the scale $\Lambda$. In the high-temperature phase, it is convenient to define the effective couplings
\begin{equation}
    \lambda_N(T) \equiv \lambda \left(1 + \frac{v_-(T)}{\Lambda_{\rm eff}}\right) \,, \,\,\, \lambda_{N'}(T) \equiv \lambda \left(1  -  \frac{v_-(T)}{\Lambda_{\rm eff}}\right) \,,
\end{equation}
with $\Lambda_{\rm eff} \equiv \frac{\lambda}{C} \Lambda$, in terms of which
\begin{equation}
    - \mathcal{L}_N = \lambda_N(T) \phi_+ N N + \lambda_{N'}(T) \phi_+ N' N' \,.
\end{equation}
Note that $\lambda$ now controls the overall size of the freeze-in production while ${\Lambda_{\rm eff}}$ controls the timing.

In order to prevent equilibration of the SM and mirror sectors with the thermal sector, we require $\lambda_N \ll 1$, which restricts $\lambda$ as well as the maximum temperature at which this effective theory remains sensible. During the broken phase $\lambda_N(T) > \lambda_{N'}(T)$, such that $\phi$ will couple preferentially to $N$. 

Taking the initial abundance of $N$ to be vanishing, the Boltzmann equation governing the evolution of the number density $n_N$ is given by
\begin{equation}\label{Boltzmann}
    \dot{n}_N + 3 H n_N \simeq (n_+^{\rm eq})^2 \expval{\sigma v} \,,
\end{equation}
where $n_+^{\rm eq} = \frac{T}{2\pi^2} M_+^2 K_2(\frac{M_+}{T})$ is the equilibrium number density for $\phi_+$ and $\expval{\sigma v}$ is the thermally-averaged annihilation cross section for the production of $N$. 
We presume the neutrinos are very heavy $M_+ < 2 M_N$ and that their direct couplings to $\phi_-$ are somewhat smaller than those to $\phi_+$, such that the dominant process contributing to their production will be the $2 \rightarrow 2$ annihilation $\phi_+ \phi_+ \rightarrow N N$. This occurs at tree level via t- and u-channel diagrams with an amplitude $|\mathcal{M}_{++ \rightarrow NN}|^2$ and cross section $\sigma_{++ \rightarrow NN}$. The thermal average appearing in Eq. (\ref{Boltzmann}) is then obtained by performing an integral over the squared center of mass energy $s$, in the manner described in \cite{Paolo:1991}. Crucially, the freeze-in rate will be proportional to $\lambda_N(T)^4$, enhancing the effect of the asymmetry in couplings. 

To solve the Boltzmann equation, it is more convenient to work with the yield $Y_N = \frac{n_N}{S}$ and reparameterize in terms of temperature via $\frac{d}{dt} \simeq - HT \frac{d}{dT}$, valid when the number of relativistic degrees of freedom in the bath remains roughly constant. Then the left-hand side becomes $\dot{n}_N + 3 H n_N = - H T S \frac{dY_N}{dT}$. Integrating, the yield as a function of temperature is
\begin{equation}\label{yield}
\begin{split}
    Y_N (T) & = \frac{2}{(4 \pi)^5} \int_{T}^{T_{\rm max}} dT' \frac{1}{H(T') S(T')}\\ 
    & \times \int_{4 M_\phi^2}^\infty ds \, \frac{\sqrt{s-4 M_N^2}\sqrt{s-4 M_+^2}}{4\sqrt{s}} K_1\left( \frac{\sqrt{s}}{T'} \right)\\
    & \times \int_{-1}^1 d\cos\theta \, |\mathcal{M}_{++ \rightarrow NN}(s,T',\theta)|^2 \,,
\end{split}
\end{equation}
where $K_1$ is a modified Bessel function of the second kind, $H = \sqrt{\frac{4 \pi^3}{45} g_\star} \frac{T^2}{M_{\rm Pl}}$, $S = \frac{2 \pi^2}{45} g_{\star, S} T^3$, and $g_\star \simeq g_{\star, S} \simeq 2$ for our scalar sector. An analogous expression holds for the $N'$ yield. The crucial difference is that the coupling $\lambda_{N'}$ becomes vanishingly small at a temperature
\begin{equation}
    T_\star \equiv \sqrt{\frac{\lambda_-}{|c_-|}} \, \Lambda_{\rm eff} \,,
\end{equation}
presuming that production takes place at temperatures far greater than $\mu_-$. Thus if the dominant freeze-in production takes place around $T_\star$, the result will be a much smaller abundance of $N'$. 

\begin{figure}[h]
\centering
\includegraphics[width=0.39\textwidth]{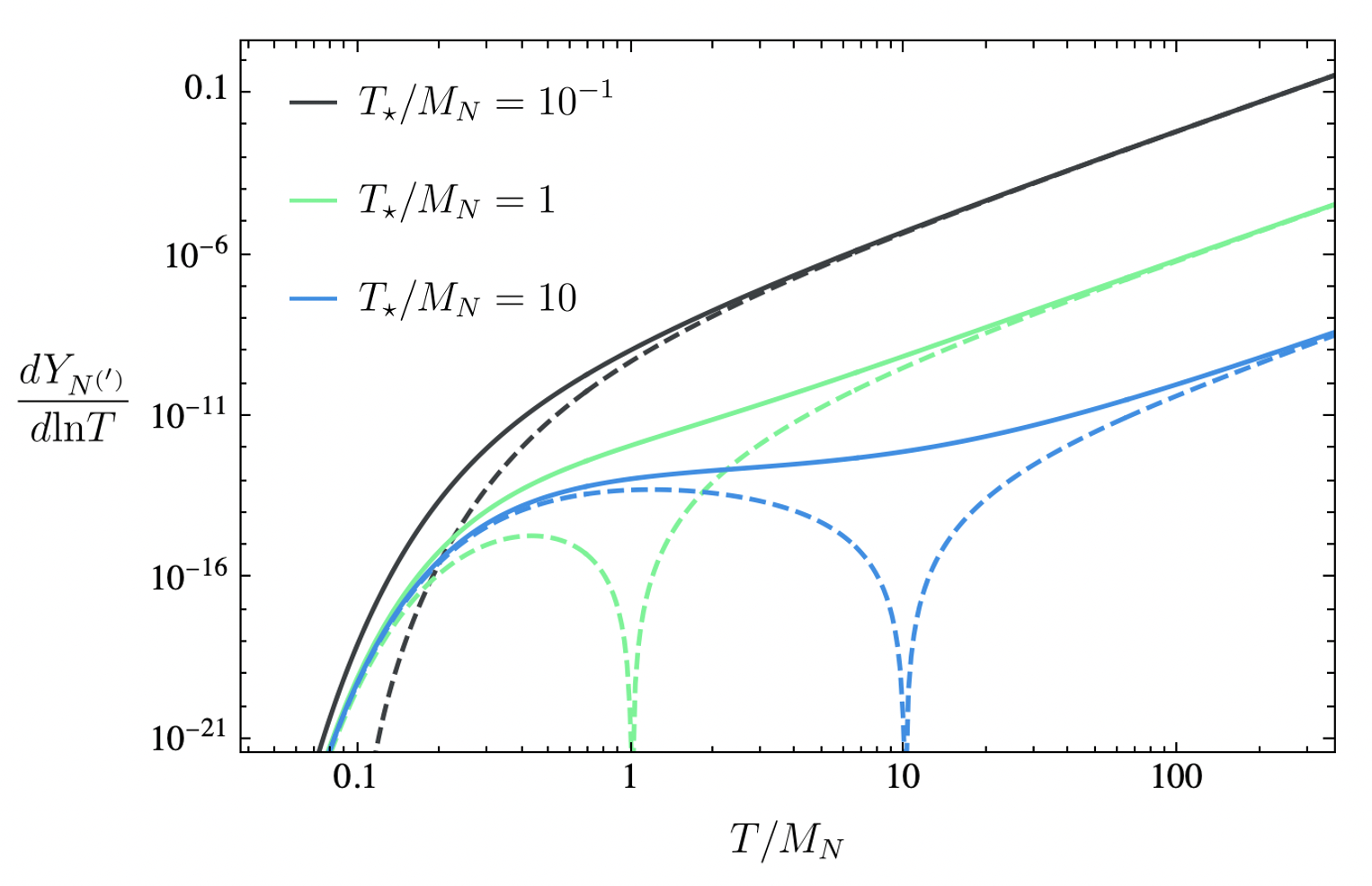}
\hspace{5mm}
\caption{Logarithmic differential yield of $N$ (solid) and $N'$ (dashed) as a function of temperature. Parameters fixed as $M_N=10^{14}$ GeV and $\lambda = 4 \times 10^{-5}$; changing either just results in an overall vertical translation.}
\label{yieldplot}
\end{figure}

What should we expect for the yield curve as a function of temperature? If $\phi_-$'s vev were temperature independent, this would be a $2 \rightarrow 2$ freeze-in through marginal operators, and so infrared dominated. The production rate would be largest at $T \sim M_N$, after which the process would become Boltzmann suppressed. This suggests $M_N \sim T_\star$ should generate appreciably asymmetric abundances. Indeed, this is observed in Fig.~\ref{yieldplot}.

However, $T_\star$ is also the temperature above which the couplings $\lambda_{N^{(')}}(T)$ become dominated by the term linear in the vev. This additional temperature-dependence results in a yield which is sensitive to the high-temperature initial conditions, as in ultraviolet freeze-in. In order for the symmetric production at high temperatures to not exceed the infrared contribution\footnote{Another potentially problematic symmetric contribution comes from the gravitational production of $N/N'$ during inflationary reheating. Graviton-mediated scattering leads to a freeze-in yield with a rate $R_{1/2} \sim T^8/M_{\rm Pl}^4$ \cite{Olive:2021,Olive:2022}, which is easily subdominant so long as $T_{\rm max}$ is not too close to the Planck scale.}, spoiling our mechanism, we require the process shut off at some $T_{\rm max}$ not much larger than $T_\star$. Note that the consideration $M_+(T) < 2M_N$ also restricts $T_{\rm max}$ to lie not too far above $M_N$, or more precisely $T_{\rm max} \lesssim \frac{2}{c_+} M_N$. These requirements are reflected in Figs.~\ref{xRHplot} and \ref{TRHplot}.

One possibility is for the scalar sector to have only begun at $T_{\rm max}$ following inflationary reheating. Alternatively, since our calculations must anyway have some $T_{\rm max} \lesssim \Lambda$ where the EFT breaks down, it is possible that the degrees of freedom at $\Lambda$ which have generated the dimension-5 operator also contribute to the effective potential at this scale. The resultant modification to the quartic couplings could cause $c_-$ to flip signs, such that at higher energies we are once again in the $\mathbb{Z}_2$-symmetric phase. With zero vev, the freeze-in contribution from higher temperatures becomes negligible. In any case, our analysis will stay agnostic to the physics of $T_{\rm max}$.

\subsection{Asymmetric Reheating}

Reheating of the SM and mirror sectors occurs via the out-of-equilibrium decays of $N$ and $N'$, respectively. As a prerequisite, we should first ensure that the massive neutrinos are sufficiently long-lived that they come to dominate over the radiation energy density in the thermal scalar bath before they decay (see Fig.~\ref{fig:timeline}). Let $R = \rho_N/\rho_\phi$ be the ratio of energy density in $N$ to that in the scalar bath. The initial value $R_*$ is set by freeze-in and quantitatively ranges from $10^{-8}$ to $10^{-12}$ for the parameter space in which we can have a successful asymmetric reheating in our toy model. Since the heavy neutrinos are non-relativistic, their energy density dilutes as $\rho_N \propto a^{-3}$ while that in the scalar bath falls as $\rho_\phi \propto a^{-4}$, meaning $R$ grows as $a \sim 1/T$.  By the time of $N$ decay, $R_{\rm dec} = T_* R_*/T_{\rm dec} \simeq M_N R_*/T_{\rm dec} $. The temperature $T_{\rm dec}$ at which $N$ decays is roughly set by $\Gamma_N \sim y_\nu^2 M_N \simeq H(T_{\rm dec})$, allowing us to identify $T_{\rm dec} \sim y_\nu \sqrt{M_N M_{\rm Pl}}$. For $N$ to dominate at decay, we require $R_{\rm dec} > 1$, corresponding to the upper bound on the Yukawa coupling
\begin{equation}\label{yukawabound1}
    y_\nu < \sqrt{\frac{M_N}{M_{\rm Pl}}} R_* \,.
\end{equation}
Thus $N$ can easily be made to dominate at decay by turning down the Yukawa coupling $y_\nu \tilde{H} L N$, which is technically natural. 

To determine the parameter space corresponding to a successful reheating, we should calculate the final ratio of temperatures $x_{\rm RH} = T_{\rm mirror}/T_{\rm SM}$, which must be sufficiently small, as well as the absolute scale of the SM reheating temperature, which should be at least $T_{\rm RH} \gtrsim 10\,\text{MeV}$ to ensure the predictions of big bang nucleosynthesis (BBN) are unaffected. In the instantaneous decay approximation, we can estimate the SM reheating temperature $T_{\rm RH}$ as
\begin{equation}\label{TRH}
    T_{\rm RH} = \left( \frac{30}{\pi^2 g_\star} \rho_{N}(t_{\rm dec}) \right)^{1/4} \,,
\end{equation}
where $g_\star$ now counts the SM degrees of freedom at $T_{\rm RH}$, and $\rho_{N}(t_{\rm dec})$ is the energy density in $N$ at their decay. 

\begin{figure}[h]
\centering
\includegraphics[width=0.37\textwidth]{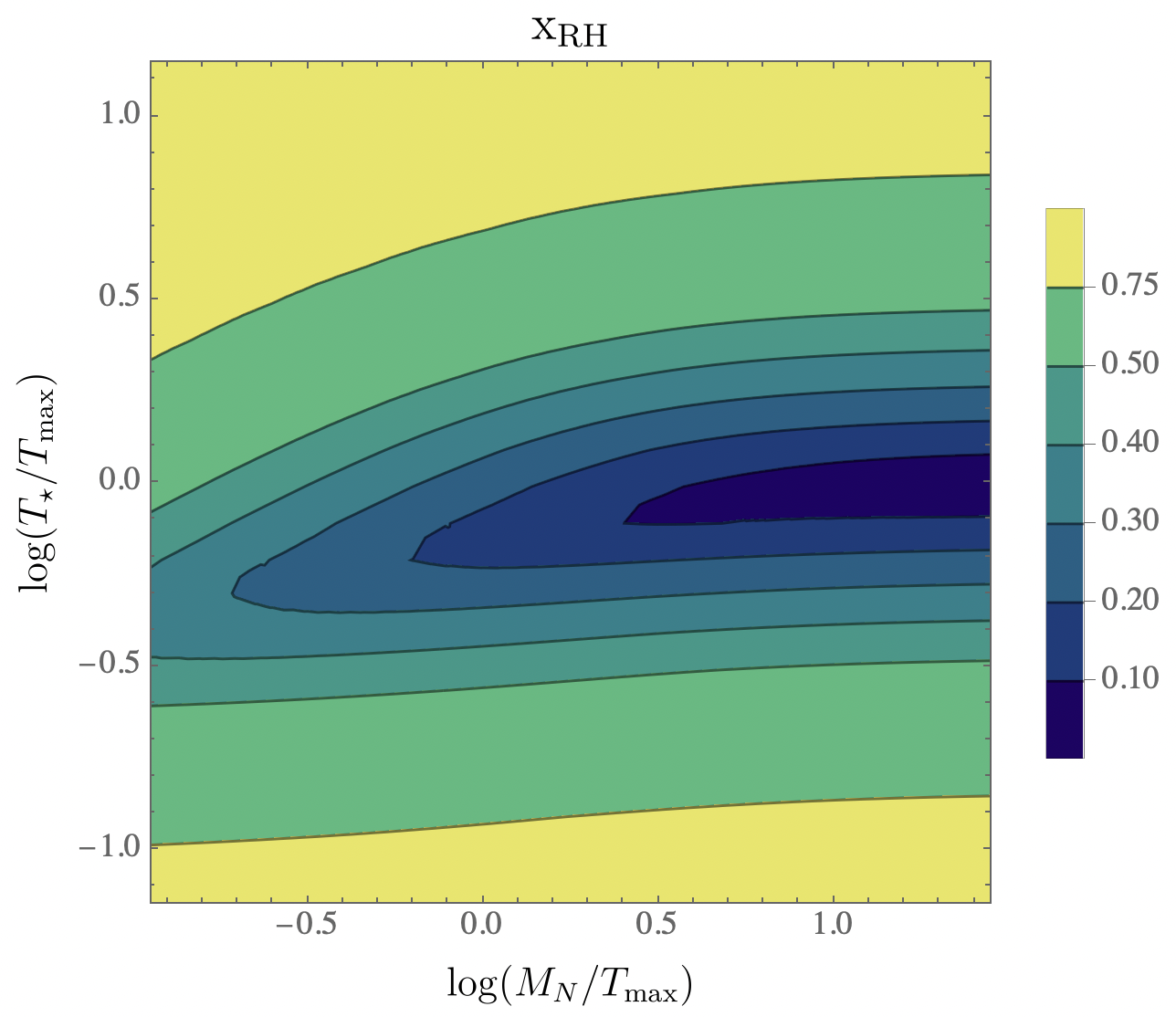}
\hspace{5mm}
\caption{The ratio of energies injected into the mirror and SM sectors, as a function of ratios of important scales. The overall energy scale is arbitrary so long as the scalar sector is in the high-temperature regime.}
\label{xRHplot}
\end{figure}

We define the ratio of energy densities
\begin{equation}
    x_{\rm RH} \equiv \left(\frac{\rho_{N'}(t_{\rm dec})}{\rho_{N}(t_{\rm dec})} \right)^{1/4} \simeq \left(\frac{Y_{N'}}{Y_{N}} \right)^{1/4}\,,
\end{equation}
and note that $x_{\rm RH}$ coincides with the final ratio of temperatures $T_{\rm mirror}/T_{\rm SM}$ at late times once the only remaining light degrees of freedom in each sector are the photon and active neutrinos, provided the asymmetry is not erased by processes which bring the SM and mirror sectors into thermal equilibrium. We have verified that the rate for scalar mediated $N$-$N'$ scattering satisfies $\Gamma = n \langle \sigma v \rangle \lesssim H$ and so is inefficient for all parameter space of interest. In Fig.~\ref{xRHplot} we plot the values of $x_{\rm RH}$ that can be realized in this toy model. 

The light species of the mirror sector contribute to the excess radiation energy density, as parameterized by the change in the effective number of neutrino species $\Delta N_{\rm eff}$, 
\begin{equation}
    \Delta N_{\rm eff} \simeq \frac{29}{7} \left( \frac{11}{4} \right)^{4/3} x_{\rm RH}^4 \,.
\end{equation}
Demanding $\Delta N_{\rm eff} \lesssim 0.5$, corresponding to the $2\sigma$ constraint from Planck \cite{Planck:2018vyg}, requires $x_{\rm RH} \lesssim 0.42$. Comparing with Fig.~\ref{xRHplot}, we see that this is indeed achievable provided $T_*$ is not too far from $T_{\rm max}$.

\begin{figure}[h]
\centering
\includegraphics[width=0.37\textwidth]{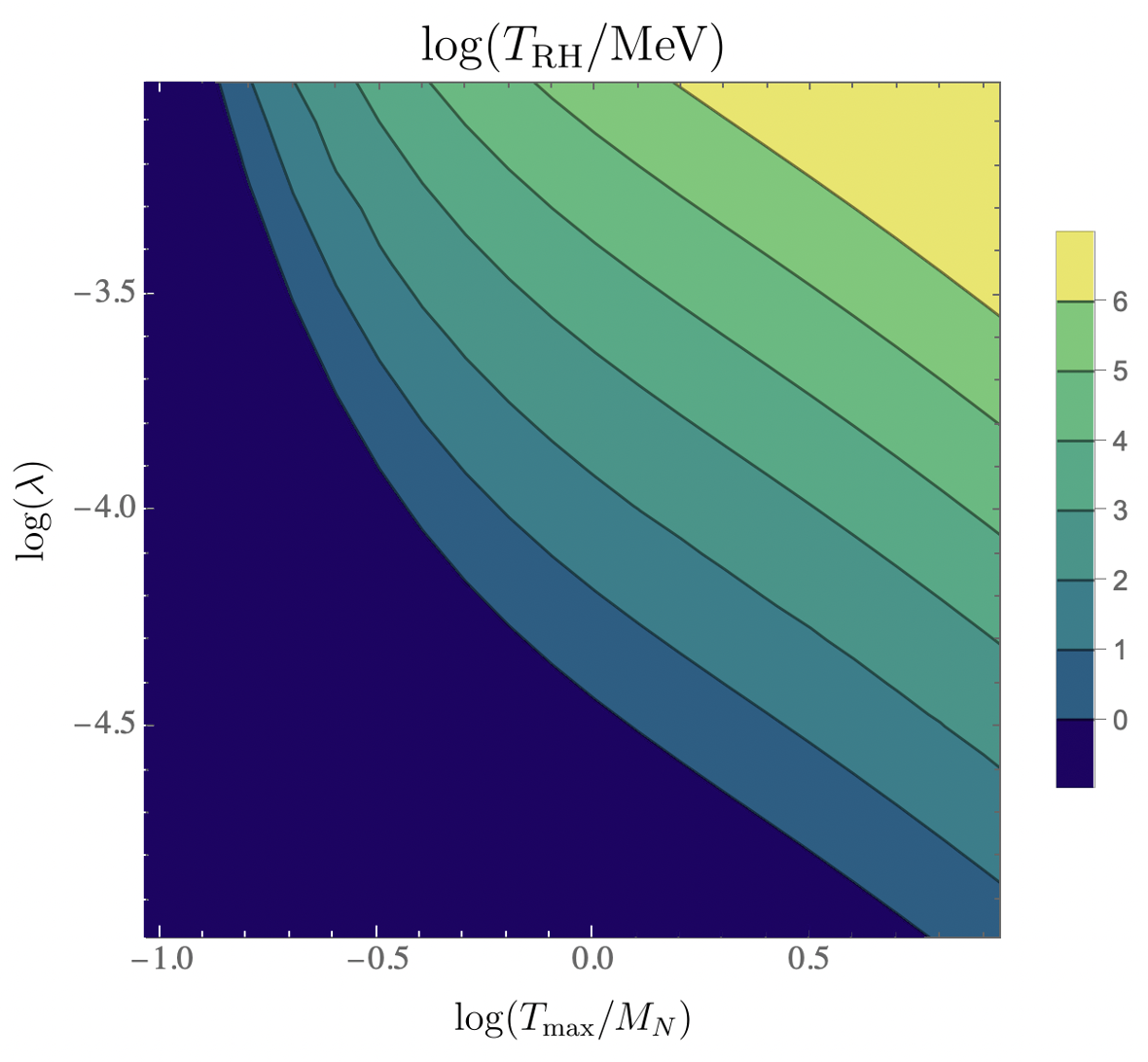}
\hspace{5mm}
\caption{Maximum reheating temperature of the SM sector, fixing $T_\star = M_N$. $T_{\rm RH}$ may be turned down by moving to $T_\star \ll M_N$ or by further increasing the neutrino lifetime.}
\label{TRHplot}
\end{figure}

To obtain the absolute scale of the reheating temperature, we must track the evolution of the energy densities from freeze-in to decay. Since we choose values for the couplings such that the sectors remain decoupled, the dominant effect governing the evolution is simply dilution due to cosmic expansion. For heavy neutrinos which decay not too long after coming to dominate the universe, we have the following approximate expression for the SM reheating temperature,
\begin{equation}
     T_{\rm RH} \simeq \frac{4}{3} \left(\frac{2}{g_\star}\right)^{\frac{1}{4}}(1+ x_{\rm RH}^4)^{\frac{3}{4}} Y_N M_N \,.
\end{equation}
Note the product $Y_N M_N$ is insensitive to the overall scale of freeze-in, since the only other scale in Eq.~(\ref{yield}) is a factor $M_{\rm Pl}$ from Hubble. The very rough estimate $T_{\rm RH}~\sim~\lambda^4 M_{\rm Pl}$ works surprising well, as observed in Fig.~\ref{TRHplot}. We require that the SM is reheated to at least $T_{\rm RH} \gtrsim 10 \ \rm{MeV}$ to ensure BBN is not substantially affected. This limits the absolute scale of the yield and prevents realizing $T_{\rm max} \ll M_N$ and freezing in solely during the Boltzmann tail, despite this still producing a large asymmetry of yields.

Finally in order to ensure a consistent late-time cosmology, we turn to the fate of the scalar sector. The leading decay channel for both $\phi$'s is to active neutrinos and proceeds through off-shell heavy neutrinos with a heavily suppressed rate $\Gamma_{\phi} \sim \lambda^2 y_\nu^4 T^5/M_N^4$ at early times and $\Gamma_{\phi} \sim \lambda^2 y_\nu^4 (v_h/M_N)^4 \mu_{\pm}$ at late times once the scalars have become non-relativistic, where $v_h$ is the Higgs vev. Crucially these rates go as $y_\nu^4$, and given the tiny values of $y_\nu$ required for the massive $N$'s to dominate the energy density prior to decay, the corresponding scalar lifetimes can easily be made significantly longer than the age of the universe. It will thus generically be the case that the scalars are stable on cosmological time scales. 

The scalars must be non-relativistic by BBN so as not to contribute to $\Delta N_{\rm eff}$, which restricts the bare masses $\mu_{\pm} \gtrsim 10\, \text{MeV}$. Being non-relativistic and stable at late times, the scalars then constitute some component of the dark matter (DM), and are harmless as long as their relic abundance is not too large. 

To check this, we define the new ratio $\tilde{R} = \rho_{\rm SM}/\rho_\phi$, and demand $\tilde{R} \gtrsim 1$ from the time the heavy neutrinos decay up through shortly after matter-radiation equality, such that $\rho_\phi$ does not come to dominate appreciably over the SM radiation bath. The assumption of instantaneous decay $\rho_N \simeq \rho_{\rm SM}$ leads to the initial condition $\tilde{R}_{\rm dec} \simeq R_{\rm dec}$. Neglecting SM entropy dumps, the ratio remains roughly fixed until the scalars become non-relativistic at some time $t_{\rm nr}$ corresponding to $T_{\rm nr} \sim \mu_\pm$, leading to $\tilde{R}_{\rm nr} \simeq \tilde{R}_{\rm dec}$. Afterwards, $\rho_\phi$ will begin to grow relative to the still-relativistic $\rho_{\rm SM}$, leading $\tilde{R}$ to decrease as $1/a \sim T$. Demanding that $\tilde{R} \gtrsim 1$ through matter-radiation equality of the SM sector imposes the constraint $R_{\rm dec} \gtrsim 10^7 (\mu_\pm/10 \,\text{MeV})$, implying
\begin{equation}
    y_\nu \lesssim 10^{-7} R_* \sqrt{\frac{M_N}{M_{\rm Pl}}} \left( \frac{10\, \text{MeV}}{\mu_\pm} \right) \,.
\end{equation}
This is a stronger condition than Eq.~(\ref{yukawabound1}), but can still be compatible with technically natural values for $y_\nu$. For example, a benchmark point for successful asymmetric reheating with $R_*=10^{-8}$ and $M_N = 10^{17}\,\text{GeV}$ would correspond to at maximum $y_\nu \sim 10^{-16}$. 

%Observationally the present day ratio of energy density in dark matter to that in baryonic matter is $\Omega_{\rm DM}/\Omega_b \simeq 5$, so in order for the scalars to constitute all of the DM we would want a final value at SM matter-radiation equality $\tilde{R}_{\rm eq} \sim 1/5$. Realistically since the scalars will constitute warm dark matter and for a large component we would conceivably have to worry about bounds from structure formation and clustering, it is best to keep them as a subdominant component, hence our more conservative bound of $\tilde{R}_{\rm eq} \gtrsim 1$.

%%%%%%%%%%%%%%%%%%%%%%%%%%%%%%%%%%%%%%%%%%%%%%%%%%%%%%%%%%%%%%%%%%%%
%%%%%%%%%%%%%%%%%%%%%%%%%%%%%%%%%%%%%%%%%%%%%%%%%%%%%%%%%%%%%%%%%%%%
\section{Conclusion}

In this work we have constructed a model of asymmetric reheating using the finite temperature phenomenon of inverse symmetry breaking. Our focus has been on constructing a minimal realization of this mechanism, which has the benefit of providing a module which may be annexed onto a variety of theories.

This reinvigorates the well-motivated scenario of degenerate mirror models and the rich phenomenology that accompanies them. A clear direction for future work is to further integrate this into such models or other theories of the early universe---perhaps exploring connections to leptogenesis or further developing the connection to dark matter.

Finally, this mechanism does require a rough confluence of scales to produce an appreciable temperature asymmetry. We emphasize that this is not an instance of fine-tuning --- to ask that the dimensionful scales in a new sector be of the same order of magnitude is exactly what one expects in a natural theory where there is some underlying scale $\Lambda$ and the relevant physics is controlled by this scale and order-one couplings. Still, it would be pleasing to study concrete models where, for example, $T_\star$ and $T_{\rm max}$ arise from the same additional degrees of freedom interacting with $\phi_+$ and $\phi_-$. 

%%%%%%%%%%%%%%%%%%%%%%%%%%%%%%%%%%%%%%%%%%%%%%%%%%%%%%%%%%%%%%%%%%%%
%%%%%%%%%%%%%%%%%%%%%%%%%%%%%%%%%%%%%%%%%%%%%%%%%%%%%%%%%%%%%%%%%%%%
\section{Acknowledgements}

We thank Gordan Krnjaic, Robert McGehee, and Carlos E.M. Wagner for feedback on an earlier version of this manuscript. We thank an anonymous referee for comments which helped improve the quality of the manuscript.
AI was supported by the U.S. Department of Energy, Office of Science, Office of Workforce Development for Teachers and Scientists, Office of Science Graduate Student Research (SCGSR) program, which is administered by the Oak Ridge Institute for Science and Education (ORISE) for the DOE. ORISE is managed by ORAU under contract number DE‐SC0014664. SK was supported by an Oehme Postdoctoral Fellowship from the Enrico Fermi Institute at the University of Chicago, and performed this work in part at the Aspen Center for Physics, which is supported by National Science Foundation grant PHY-1607611.

\bibliography{inverse}

\end{document}